\documentclass[showpacs,preprintnumbers,pre,twocolumn]{revtex4}
\usepackage{amsmath,amssymb,graphicx,subfigure,color}

\begin{document}
\title{Phase space gaps and ergodicity breaking in systems
with long range interactions}
\author{
Freddy Bouchet$^{1,2}$\thanks{E-mail: Freddy.Bouchet@inln.cnrs.fr},
Thierry Dauxois$^1$\thanks{E-mail: Thierry.Dauxois@ens-lyon.fr},
David Mukamel$^{1,3}$\thanks{E-mail: }, Stefano
Ruffo$^{1,4}$\thanks{E-mail: stefano.ruffo@unifi.it}} \affiliation{
1. Universit\'e de Lyon, CNRS, \'Ecole Normale Sup\'erieure de Lyon
Laboratoire de Physique, 46 All\'{e}e d'Italie, 69364 Lyon cedex 07,
France\\
2. Institut Non Lin\'eaire de
Nice, CNRS, 1361 route des Lucioles 06560 Valbonne, France.\\
3. Department of Physics of
Complex Systems, The Weizmann Institute of Science, Rehovot, Israel \\
4. Dipartimento di Energetica ``S. Stecco" and CSDC, Universit{\`a}
di Firenze and INFN, Via S. Marta, 3 I-50139, Firenze, Italy}

\date{\today}

\begin{abstract}
We study a generalized isotropic XY-model which includes both
two-spin and four-spin mean-field interactions. This model can be
solved in the microcanonical ensemble. It is shown that in certain
parameter regions the model exhibits gaps in the magnetization at
fixed energy, resulting in ergodicity breaking. This phenomenon
has previously been reported in anisotropic and discrete spin
models. The entropy of the model is calculated and the
microcanonical phase diagram is derived, showing the existence of
first order phase transitions from the ferromagnetic to a
paramagnetic disordered phase. It is found that ergodicity
breaking takes place both in the ferromagnetic and the
paramagnetic phases. As a consequence, the system can exhibit a
stable ferromagnetic phase within the paramagnetic region, and
conversely a disordered phase within the magnetically ordered
region.
\end{abstract}
\pacs{05.70.Fh Phase transitions: general studies}
 \maketitle

\section{Introduction}

The aim of this paper is to shed light on a dynamical feature
which distinguishes systems with long range interactions from
those with short range ones. It is well known that the attainable
region in the space of extensive thermodynamic parameters is
always convex when only short range interactions are present.
Consider, for example, two magnetic subsystems with the same
potential energy~$V$ but with two different magnetization values
$m_1$ and $m_2$ (notice that, instead of the magnetization, we
might have chosen any other order parameter). Introducing a
parameter $\lambda$ which depends on the relative size of the
subsystems, it is possible to get any intermediate value of the
magnetization~$m$ by just combining the two subsystems with
appropriate weights, such that $m=\lambda m_1+(1-\lambda)m_2$,
while the potential energy is kept constant to $V$. It is
important to stress that this convexity property is satisfied only
if the system is additive, since the interaction energy between
the two subsystems has been neglected. This can safely be done
only for short range interacting systems in the large volume
limit. Moreover, convexity implies that the accessible region in
the space of thermodynamic parameters is connected.

In contrast, systems with long range interactions \cite{Houches}
are not additive. As a consequence, the convexity property may be
violated, and then also connectivity. This feature has profound
consequences on the dynamics. Gaps may open up in the order
parameters space. Such gaps have been recently reported in a class
of anisotropic $XY$ models~\cite{borgo,borgobis} and for a
discrete spin system~\cite{schreiber}. Since the accessible region
in the order parameters space is no more connected, ergodicity
breaking naturally appears when a continuous microcanonical
dynamics is considered. This is the property we discuss in this
paper. More specifically, the question we would like to address is
that of locating the region in the phase diagram of the model
where we expect ergodicity breaking to occur. We will argue that
ergodicity breaking occurs in regions of parameters where
metastable states exist. In such regions, one may also generically
expect first order phase transitions. Thus, we conclude that the
breakdown of order parameters space connectivity is generically
associated with first order phase transitions. Moreover, we
conjecture that the intersection between first order phase
transition lines and the boundary of the parameter region where
the order parameters space is disconnected, occurs at non smooth
(singular) points of the boundary itself. We also conjecture that
the same is true for the metastability line.

In recent years, models describing classical rotators with
all-to-all mean-field interaction have been intensively studied in
order to explain general dynamical features of long-range
interacting systems~\cite{Houches}. The most celebrated of such
models is the so-called Hamiltonian Mean Field (HMF) model, where
rotators denoted by the angle $\theta_i$, $i=1,\dots,N$ are
coupled through a long-range mean-field potential
\begin{equation}
W=-\frac{J}{2N} \sum_{i,j} \cos (\theta_i-\theta_j).
\label{potential}
\end{equation}
The interaction is ferromagnetic for positive $J$-values, and
antiferromagnetic otherwise. Defining the complex order parameter
(with $i$ the imaginary unit)
\begin{equation}
\overrightarrow{m}=(m_x+im_y)= \frac{1}{N} \sum_n \exp (i \theta_n),
\label{order}
\end{equation}
the potential per particle $V=W/N$ can be rewritten as
${-Jm^2}/{2}$ where $m=|\overrightarrow{m}|$ is the modulus of the
magnetization. The total energy is finally obtained by adding a
kinetic energy term to the above potential so that we get the
Mean-Field Hamiltonian
\begin{equation}
H=\sum_{n=1}^N \frac{p_n^2}{2}+N\,V.
\end{equation}
In this formulation, the model can also be thought of as
describing a system of particles with unit mass, interacting
through the mean-field coupling (the names rotator and particle
might equivalently be used). Note that a system with non-unit mass
may be reduced to the above system by rescaling time. As the
parameter $J$ might be removed by a suitable energy rescaling, we
will omit it in the remainder of the paper. Within the Hamiltonian
framework, both the energy density $E=H/N$ and the total momentum
$P= \sum_n p_n$ are conserved quantities and they are fixed by the
initial conditions.

In this paper, we study a generalized mean-field
Hamiltonian~\cite{debuyl} with potential
\begin{equation}
V\left( m\right) =\frac{m^{2}}{2}-K\frac{m^{4}}{4},
\label{hamiltonianmdeuxmquatre}
\end{equation}
which reduces to the antiferromagnetic HMF model ($J=-1$) in the particular
case $K=0$. In addition to the first antiferromagnetic term, we
consider a ferromagnetic fourth order term in the magnetization,
whose intensity is controlled by a {\em positive} constant $K$. In
the regime of large values of the parameter~$K$, a magnetized
state would be favorable because of the ferromagnetic coupling
while, for small values of $K$, the antiferromagnetic coupling
dominates and leads to a non-magnetic state. As we shall
demonstrate below, there exists a range of the parameter $K$ for
which the model exhibits a first order phase transition between a
paramagnetic phase at high energies and a ferromagnetic phase at
low energies. In both phases there are regions in the $(E,K)$
plane in which the accessible magnetization interval exhibits a
gap, resulting in breaking of ergodicity.

The organization of the paper is as follows.  In
Section~\ref{phasediagram}, we first address the issue of the
accessible magnetization interval in this model. In the following
Section~\ref{mecastat}, we present the statistical mechanics of the
model. In particular, we study the phase diagram and the conditions
for phase transitions to occur. In Section~\ref{dynamics}, we study
the dynamical consequences of this ergodicity breaking. Finally,
Section~\ref{conclusion} will be devoted to conclusions and
perspectives.

\section{Phase space connectivity}
\label{phasediagram}

Let us study the phase space structure and the phase diagram of this
model. The specific kinetic energy $E_K=E-V\left( m\right)$ is by
definition a non negative quantity, which implies that
\begin{equation}
\label{inequality}E\geq V(m)=m^{2}/2-Km^{4}/4.
\end{equation}
We will show that as a result of this condition not all the values of
the magnetization $m$ are attainable in a certain region in the
$(E,K)$ plane; a disconnected magnetization domain might be indeed
a typical case. As  explained below, this situation is the one of
interest. Let us characterize the accessible domains in the
$(E,K)$ plane more precisely  by analyzing the different regions.

\begin{figure}[htbp]
\centering\includegraphics[width=7cm]{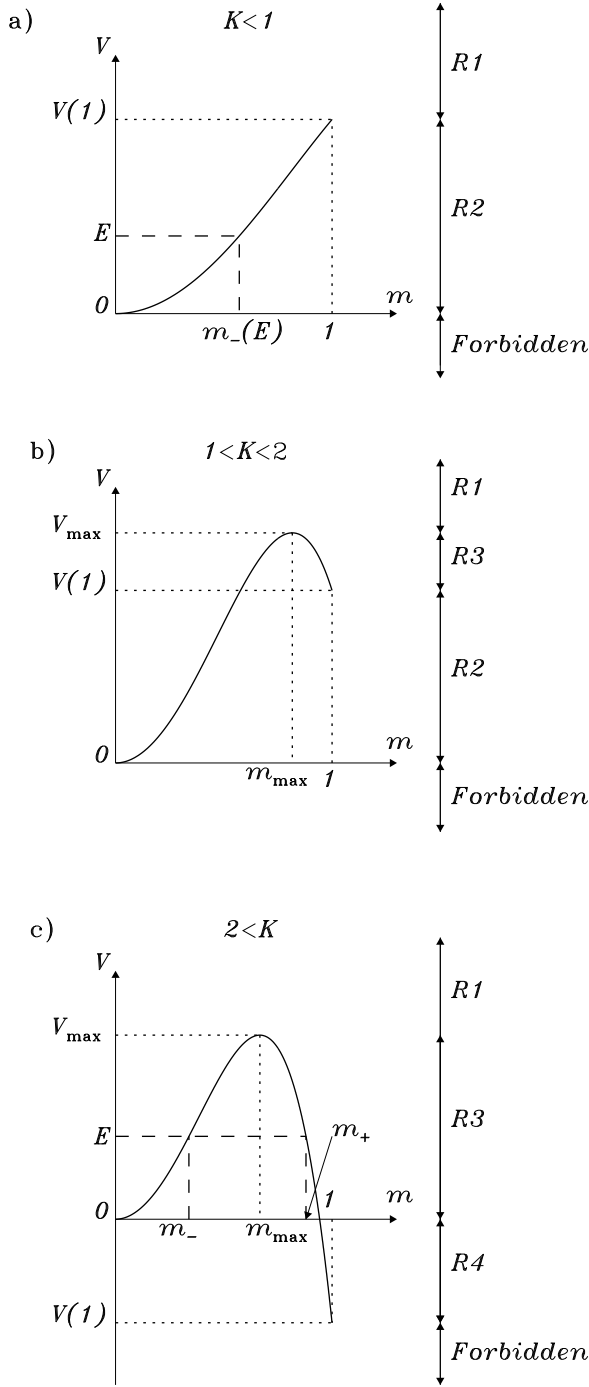}
\bigskip
\caption{Specific potential energy $V$ vs. magnetization $m$ for
three different cases:  $K<1$ (panel a), $1<K<2$ (panel b) and
$2<K$ (panel c). The location of the maximal magnetization
$m_{\max}$ and the corresponding potential energy $V_{\max}$ are
shown (see text). In panels (a) and (c), two examples of the
location of the critical magnetization $m_\pm(E,K)$ is indicated
for energy values $E$ in the intermediate regions.
}\label{fig:diffmagnetization}
\end{figure}

\begin{figure}[htbp]
\includegraphics[width=9cm]{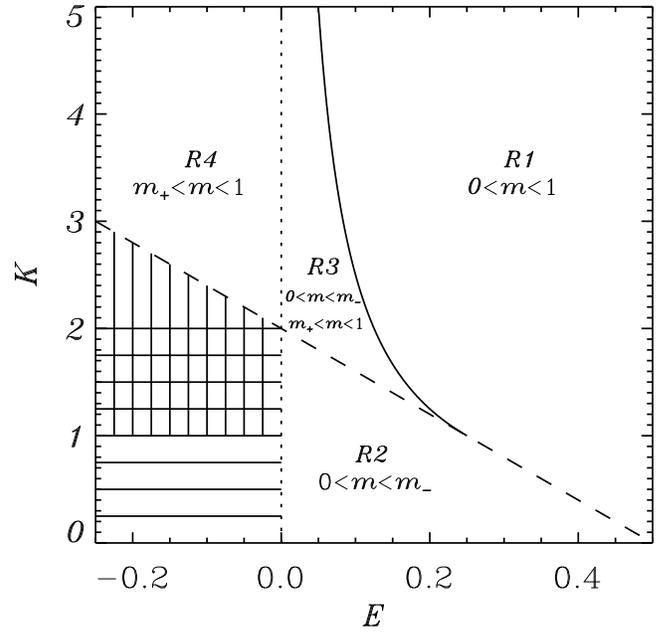}
\caption{The $(E,K)$ plane is divided in several regions. The
solid curve corresponds to $K=1/(4E)$, the oblique dashed line to
$K=2-4E$, while the dotted one to $K=1$. The vertically shaded, quadrilled,
and horizontally shaded regions are forbidden.
The accessible magnetization interval in each of the four regions
is indicated (see text for details).}\label{fig:energy_range}
\end{figure}

For $K<1$, the local maximum of the potential energy $V$ is not
located inside the magnetization interval $[0,1]$ (see
Fig.~\ref{fig:diffmagnetization}a). The potential being a strictly
increasing function of the magnetization, the maximum is reached
at the extremum $m=1$. The complete interval $[0,1]$ is thus
accessible, provided the energy $E$ is larger than $V(1)$: the
corresponding domain is in {\em R1} defined in
Fig.~\ref{fig:energy_range}. The horizontally shaded region is
forbidden, since the energy is lower than the minimum of the
potential energy $V(0)=0$. Finally, the intermediate region
$0<E<V(1)$ in included in the region~{\em R2}: it is important to
emphasize that only the interval $\left[0,m_{-}\left(
E,K\right)\right]$, where $m_{\pm }\left( E,K\right) =[({1\pm
\sqrt{1-4EK}})/{K}]^{1/2}$, is accessible. Larger magnetization
values correspond to a potential energy~$V(m)$ larger than the
total energy~$E$, which is impossible.
Figure~\ref{fig:diffmagnetization}a also displays the value $m_-$
corresponding to an energy  in the intermediate region {\em R2}.

For $K\geq1$, the specific potential energy $V$ has a maximum
$V_{\max }=1/4K$ which is reached at $0<m_{\max
}=1/\sqrt{K}\leq1$. Figures~\ref{fig:diffmagnetization}b and
~\ref{fig:diffmagnetization}c, where the potential
energy~(\ref{hamiltonianmdeuxmquatre}) is plotted {\em vs} the
magnetization $m$, present such  cases. For an energy $E$ larger than
the critical value $V_{\max }$, condition ~(\ref{inequality})
is satisfied for any value of the magnetization $m$. The complete
interval $\left[ 0,1\right]$ is thus accessible for the
magnetization $m$. This  region is {\em R1} represented
in Fig.~\ref{fig:energy_range}.

Let us now consider the cases for which $E\leq V_{\max}$. As
discussed above, the minimum $V_{\min}$ of the potential energy is
also important to distinguish between the different regions. For
$1\leq K\leq2$ (see Fig.~\ref{fig:diffmagnetization}b), the
minimum of $V(m)$ corresponds to the non-magnetic phase $m=0$
where $V(0)=0$. The quadrilled region shown in
Fig.~\ref{fig:energy_range}, which corresponds to negative energy
values, is thus non-accessible. On the contrary, positive energy
values are possible and correspond to very interesting cases,
since only sub-intervals of the complete magnetization interval [0,1]
are accessible. There are however two different cases:
\begin{itemize}
\item for $0<V(1)<E<V_{\max }$, the domain of possible
magnetization is $\left[0,m_{-}\left( E,K\right)\right] \cup\left[
m_{+}\left( E,K\right) ,1 \right]$. The above conditions are
satisfied in {\em R3} of Fig.~\ref{fig:energy_range}.

\item for $0<E<V(1)$, only the interval $\left[0,
m_{-}\left( E,K\right)\right]$ satisfies
condition~(\ref{inequality}). This takes place within {\em R2} of
Fig.~\ref{fig:energy_range}.
\end{itemize}

For  the domain $2\leq K$, the minimum of the potential energy is
attained at the extremum, $m=1$, implying $E>V(1)=1/2-K/4$. The
vertically shaded region is thus forbidden. In the accessible
region two cases can be identified:

\begin{itemize}
\item for $V(1)<E<0$, only the interval $\left[ m_{+}\left(
E,K\right) ,1\right] $ satisfies condition~(\ref{inequality}). It
is important to note that $m_{+}\left( E,K\right) \leq 1 $
provided $E\geq 1/2-K/4 $. These cases correspond to {\em R4}.

\item for $0\leq E\leq V_{\max}$, the two intervals
$\left[0, m_{-}\left( E,K\right)\right] $ and $ \left[ m_{+}\left(
E,K\right) ,1 \right]$ satisfy condition~(\ref{inequality}),
corresponding to {\em R3} of Fig.~\ref{fig:energy_range}.
\end{itemize}

\begin{figure}[htbp]
\centering\includegraphics[width=7cm]{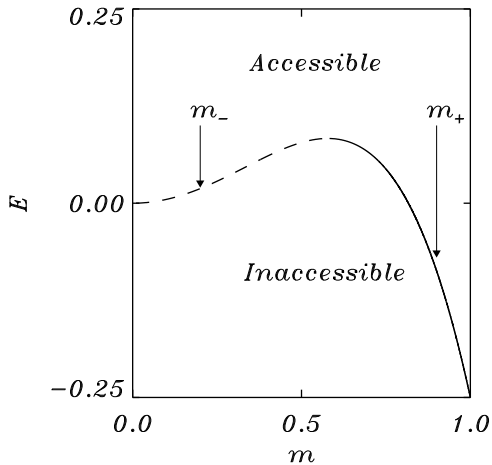}
\caption{Accessible region in the $(m,E)$ plane for $K=3$. For
energies in a certain range, a gap in the accessible magnetization
values is present and defined by the two boundaries $m_\pm(E,K)$.
}\label{fig:limitelagnetization}
\end{figure}

In summary, the complete magnetization interval  $[0,1]$ is
accessible only in the region{\em R1}. In {\em R2}, only
$\left[0,m_{-}\right]$ is accessible, while only
$\left[m_{+},1\right]$ is accessible in {\em R4}. Finally, we note
that the phase space of the system is not connected in the region
{\em R3}. Indeed, the magnetization cannot vary continuously from
the first interval $\left[0,m_{-}\right]$ to the second one
$\left[m_{+},1\right]$, although both are accessible. These
restrictions yield the accessible magnetization domain shown in
Fig.~\ref{fig:limitelagnetization}. The fact that for a given
energy the phase space is disconnected implies ergodicity breaking
for the Hamiltonian dynamics. It is important to emphasize that
the discussion above is independent of the number of particles and
ergodicity is expected to be broken even for a finite (but sufficiently
large) $N$.

\section{Statistical mechanics}
\label{mecastat}

We have thus found that in certain regions in the $(E,K)$ plane,
the magnetization cannot assume any value in the interval
$[-1,1]$. For a given energy there exists a gap in this interval
to which no microscopic configuration can be associated. In this
Section, we study the statistical mechanics of model
(\ref{hamiltonianmdeuxmquatre}), by considering the microcanonical measure
\begin{equation}
\mu(E,N) =\prod ^{N}_{k=1}\mbox{d} \theta _{k}\mbox{d} p_{k}\,
\delta \left( H-NE\right).
\end{equation}
The probability $P(m)$ to observe the magnetization $m$ can be
obtained for large values of the number~$N$ of particles using
large deviation techniques, in a similar way as it was derived for
the HMF model~\cite{bbdr}. Here again,  the appropriate global
variables are the magnetization $m$ and the mean kinetic energy
$E_K=\sum_n p_n^2/N$. Cram\'er's theorem allows one to derive the probability
distribution $P(m)$ in the large $N$-limit. The entropy per
particle, for given energy and magnetization, is obtained as
\begin{equation}
s(E,m)=\lim_{N\rightarrow\infty}\frac{1}{N}\ln P(m),
\end{equation}
Following the steps described in Ref.~\cite{bbdr}, one finds that
the specific entropy $s$ can be expressed as
\begin{equation}
s\left( E,m\right) =s_K(E,m)+s_{conf}\left( m\right),
\end{equation} {\em i.e.} the sum of the
momentum entropy, which is related to the kinetic energy
\begin{eqnarray}
s_K(E,m)=\ln \left(E_{c}\right) /2 =\frac{1}{2}\ln \left(
E-\frac{m^{2}}{2}+K\frac{m^{4}}{4}\right),\label{eqentrokin}
\end{eqnarray}
and of the configuration entropy which, as for the
HMF model, is
\begin{equation}
s_{conf}\left( m\right) =-m\phi (m)+\ln \left( I_{0}\left( \phi
(m)\right) \right).
\end{equation}
Here $\phi $ is the inverse of the function $I_{1}/I_{0}$ where
$I_{n}$ is the modified Bessel function of order $n$. The
microcanonical thermodynamics is finally recovered solving the
variational problem
\begin{equation}
S(E)=\sup_m\left[ s(E,m)\right].
\end{equation}

\begin{figure}[htbp]
\centering\includegraphics[width=8cm]{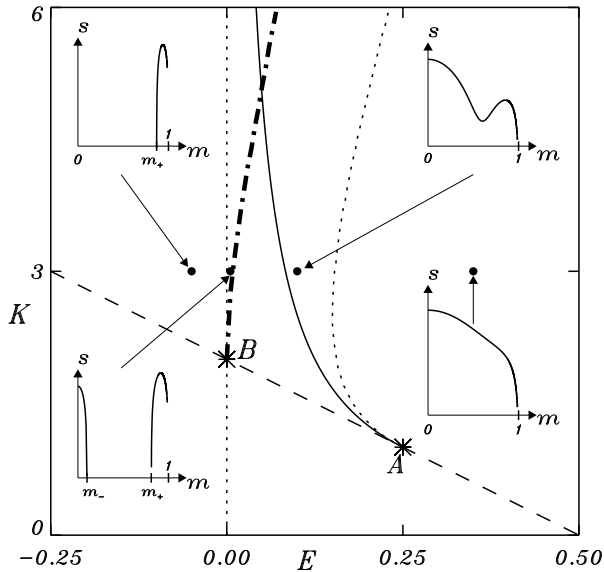} \caption{Phase
diagram of the mean field model (\ref{hamiltonianmdeuxmquatre}).
The dash-dotted curve corresponds
to the first order phase transition line, issued from the point
$B(0,2)$. As in Fig.~\ref{fig:energy_range}, the solid curve
indicates  the right border of the region {\em R3}, where the
phase space is disconnected. The dashed line corresponds to the
$K=2-4E$. The dotted line issued from the point $A(1/4,1)$
represents the metastability line for the magnetized state, while
the $E=0$ vertical dotted line is also the metastability line for the non-magnetic
state. The four insets represent the entropy $s$ versus the
magnetization $m$ for the four energies: $E=$-0.05,
0.005, 0.1 and 0.35, when $K=3$.} \label{fig:phase_diagram}
\end{figure}

First, by comparing the low and high energy regimes, it is possible
to show that a phase transition is present between the two regimes.
In the domain {\em R4} of Fig.~\ref{fig:energy_range}, for very low
energy $E$ (close to the limiting value $1/2-K/4$), the accessible
range for $m$ is a small interval located close to $m=1$ (see
Fig.~\ref{fig:diffmagnetization}c). The maximum of the entropy $s$
corresponds therefore to a magnetized state, $m_{eq}$, located very
close to $m=1$ (see the top-left inset in Fig.~\ref{fig:phase_diagram}).
On the contrary, in the very large energy domain, the variations of
entropy  with respect to $m$ are dominated by the variations of the
configurational entropy since the kinetic entropy (see
Eq.~(\ref{eqentrokin}) with $m$ of order one and a very large energy
$E$) is roughly a constant around $s(E,0)=(\ln E)/2$. As expected,
the configurational entropy is  a decreasing function of the
magnetization: the number of microstates corresponding to a
non-magnetic macrostate being much larger that the same number for a
magnetized state. The configuration entropy has therefore a single
maximum located at $m_{eq}=0$. A phase transition takes place
between the non-magnetic state at large energy, and a magnetic state
at small energy. Moreover, as the non-magnetic state is possible
only for positive energies $E$ (see
Figs.~\ref{fig:diffmagnetization}), the transition line is located
in the domain $E\geq 0$. In this region, the quantity
$\partial_{m}^{2}s(E,0)=-(1+1/(2E))$ is negative, which ensures that,
for any value of $E$ and $K$, the non-magnetic state $m=0$ is a
local entropy maximum. The latter argument allows to exclude a
second order phase transition at a positive critical energy, since
the second derivative $\partial_{m}^{2}s(E,0)$ would have to vanish,
which is impossible. The above argument leads to the conclusion that
the phase transition must be {\em first order}.

Let us now focus on the behavior of the entropy in the region {\em
R3}, where the accessible range for $m$ is the union of two
disconnected intervals $\left[ 0,m_{-} \right] \cup \left[ m_{+}
,1\right]$. As discussed above, the total entropy $s(E,m)$ has a
local maximum in the first interval $\left[ 0,m_{-} \right] $
located at $m=0$ and associated with the entropy $s_{\max
}^{1}=s(E,0)=\log (E)/2$. In the second interval $\left[ m_{+}
,1\right]$, a maximum is also present with $s_{\max
}^{2}=s(E,m_{eq})$ where $m_{eq}\geq m_+>0$. As $s^{1}_{\max }(E)$
diverges to $-\infty$ when $E$ tends to 0, a magnetized state is
expected on the line $E=0$, as long as $s_{\max }^{2}$ remains
finite. Since $K=2$ is the only value for which $s_{\max }^{2}(0,K)$
diverges, the first order transition line originates at the point
$B(0,2)$ in Fig.~\ref{fig:phase_diagram}.
Although it is possible to study analytically the
asymptotic behavior of the transition line near this point, we can
rather easily compute numerically the location of the first order
transition line, represented by the dash-dotted line in
Fig.~\ref{fig:phase_diagram}.

Figure~\ref{fig:phase_diagram} also shows the metastability
line (the dotted line starting at point $A(1/4,1)$), for the magnetized
state $m_{eq}\neq 0$. To the right of this metastability line,
there is no metastable state (local entropy maximum for any $m>0$,
see bottom-right inset in Fig.~\ref{fig:phase_diagram}) while a
metastable state (local maximum) exists at some non vanishing
magnetization on the other side (see top-right inset in
Fig.~\ref{fig:phase_diagram}). Finally, the vertical dotted line
$E=0$ corresponds to the metastability line of the non-magnetic
state $m=0$.

\begin{figure}[htbp]
\centering{\includegraphics[width=8cm]{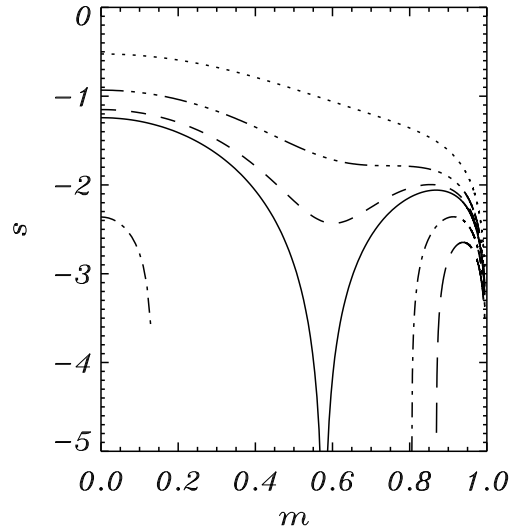}}
\caption{Entropy $s$ versus magnetization $m$ for
$K=3$ and several energy values. The different curves correspond,
from top to bottom, to $E=0.35$ (dotted), 0.155 (dash-triple
dotted, metastability limit for the magnetized state), 0.1
(dashed), $E=1/(4K)=1/12$ (solid, appearence of the gap), 0.0089
(dash-dotted, first order phase transition), -0.05 (long dashed).
This picture demonstrates that gaps in the accessible states
develop as the energy is lowered.} \label{fig:superposentropie}
\end{figure}

One of the key issues we would like to address is the possible
links between the breakdown of phase space connectivity and thus
ergodicity breaking, on the one hand, and the phase transition, on
the other hand.
Obvious general properties do exist: a region of parameters where
the phase space is disconnected corresponds to a region where
metastable states do exist.   Let us justify this statement.  At
the boundary of any connected domain,  when the order parameter is
close to its boundary value $m_b$, there is a single accessible
state. In a model with continuous variables, like the one we study
in this paper, this leads to a divergence of the entropy. In this
case the singularity of the entropy is proportional to $\ln(m-m_b)$ (see
for instance equation (\ref{eqentrokin})). For a model with
discrete variables, like an Ising model, the entropy would no more
reach $-\infty$ as $m$ tends to $m_b$, but would rather take a finite
value. However, the singularity would still exist and would then be
proportional to $(m-m_b)\ln(m-m_b)$. In both cases, of discrete and
continuous variables, at the boundary of any connected domain, the
derivative of the entropy as a function of the order parameter
tends generally to $\pm\infty$. As a consequence, entropy
extrema cannot be located at the boundary. Thus, a local entropy maximum
(metastable or stable) does exist in a region of parameters where
the phase space is disconnected.

Hence, there is an entropy maximum (either local or global) in any
connected domain of the phase space. For instance, considering the
present model, in Fig.~\ref{fig:phase_diagram}, the area {\em R3}
is included in the area where metastable states exist (bounded by
the two dotted lines and the dashed line).
In such  areas where metastable states exist, one generically
expects first order phase transitions. Thus the breakdown of phase
space connectivity will be generically associated to first order
phase transition, as exemplified by the present study. However, this
is not necessary, one may observe metastable states without first
order phase transitions, or first order phase transitions without
connectivity breaking.

A very interesting question is related to the critical points $A$
and $B$ shown in Fig.~\ref{fig:phase_diagram}. As observed in the
phase diagram, the end point for the line of first order phase
transition (point $B$)
corresponds also to a point where the boundary of the region where
the phase space is disconnected is not smooth. Similarly, the end
point for the line of appearance of metastable states (point $A$)
is also a singular point for
the boundary of the area where the phase space is disconnected. It
is thus possible to propose the conjecture that such a relation is
generic, and that it should be observed in other systems where
both first order phase transitions and phase space ergodicity
breaking do occur.

\section{Dynamics}
\label{dynamics}

The feature of disconnected accessible magnetization intervals,
which is typical of systems with long-range interactions, has
profound implications on the dynamics. In particular, starting
from an initial condition which lies within one of these
intervals, local dynamics is unable to move the system to a
different accessible interval. A macrostate change would be
necessary to carry the system from one interval to the other. Thus
the ergodicity is broken in the microcanonical dynamics even at
finite $N$.

\begin{figure}[htbp]
\centering\includegraphics[width=8cm]{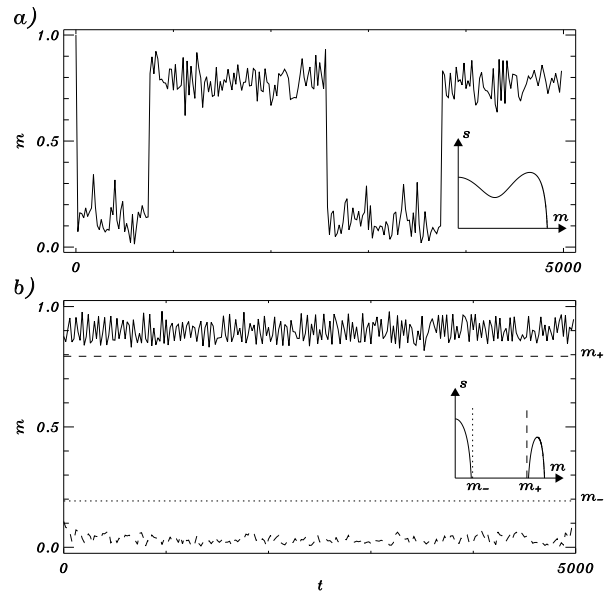} \caption{Time
evolution of the magnetization $m$ (the entropy of the
corresponding cases is plotted as an inset). Panel a) corresponds to
the case $E=0.1$ and $K=8$, while panel b) to $E=0.0177$ and $K=3$.
In panel b), two different initial conditions are plotted
simultaneously: the solid line corresponds to $m(t=0)=0.1$ while the
dashed line to $m(t=0)=0.98$. The dashed (resp. dotted) line in
panel b) corresponds to the line $m=m_+\simeq0.794$ (resp.
$m=m_-\simeq0.192$).} \label{fig:tracem}
\end{figure}

In Ref.~\cite{schreiber} this point has been demonstrated using
the microcanonical Monte-Carlo dynamics suggested by Creutz~\cite{creutz}.
Here, we use the Hamiltonian dynamics given by the equation of motions
\begin{eqnarray}
\dot p_n&=&-\frac{\partial H}{\partial
\theta_n}=-N\left(m\frac{\partial m}{\partial
\theta_n}-Km^3\frac{\partial m}{\partial \theta_n}\right)\\
&=&N\left(1-Km^2\right)\left(\sin\theta_nm_x-\cos\theta_nm_y\right).
\end{eqnarray}
We display in Fig.~\ref{fig:tracem} the evolution of the
magnetization for two cases, since we have shown above that the
gap opens up when $E$ decreases. The first case corresponds to the
domain {\em R1}, in which the
accessible magnetization domain is the full interval
$[0,1]$. Fig.~\ref{fig:tracem}a presents the time evolution of the
modulus $m$ of the order parameter~(\ref{order}) versus time. The
magnetization switches between the metastable state $m=0$ and the
stable one $m_{eq}>0$. This is possible because the number of
particles is small ($N=20$) and, as a consequence, the entropy
barrier (see the inset) can be overcome. Considering a system with a
small number of particles allows to observe flips between
local maxima, while such flips would be less frequent for larger
$N$ values.

In the other case, we consider a stable $m=0$ state which is
disconnected from the metastable one. It makes the system unable
to switch from one state to the other. Note that this feature is
characteristic of the microcanonical dynamics, since an algorithm
reproducing the canonical dynamics would allow the crossing of the
forbidden region (by moving to higher energy states, which is
impossible in the microcanonical ensemble). The result of two different
numerical simulations is reported in
Fig.~\ref{fig:tracem}b. One is initialized with a magnetization
within $[0,m_-]$, while the other corresponds to an initial
magnetization close to $m=1$ ({\em i.e.} within $[m_+,1]$). One
clearly sees that the dynamics is blocked in one of the two
possible regions, and not a single jump is visible over a long
time span. This is a clear evidence of  ergodicity breaking.

\section{Conclusions}
\label{conclusion}

We have found that for sufficiently low energy, gaps open up in the
magnetization interval, to which no microscopic configurations
corresponds. Thus the phase space breaks into disconnected regions.
Within the microcanonical dynamics the system is trapped in one
of these regions, leading to a breakdown of ergodicity even in
finite systems.


Ergodicity breaking implies that an ensemble average will not be
well reproduced by a time average, usually simpler to get from the
experimental or numerical point of view. Such a discrepancy,
forbidden {\em a priori} for short-range interacting system, might
be a typical case for systems with long-range interactions.

Of high interest would be of course to study systems with both
short and long-range interactions since they usually compete in
standard condensed matter systems \cite{CampaPRB}. Such a study could suggest an
experimental system where ergodicity breaking might be observed.

\acknowledgements This work was supported by the ANR program
STATFLOW (ANR-06-JCJC-0037-01) and by the Minerva Foundation with
funding from the Federal Ministry for Education and Research. Visit
of F. B. to the Weizmann Institute has been supported by the Albert
Einstein Minerva Center for Theoretical Physics. This work is also
funded from the PRIN05 grant of MUR-Italy {\it Dynamics and
thermodynamics of systems with long-range interactions}.

\end{document}